\documentclass{moriond}

\usepackage{xspace}
\usepackage{amsmath}
\usepackage{lineno}
\def\Journal#1#2#3#4{{#1} {\bf #2}, #3 (#4)}

\def\PRD{{\em Phys. Rev.} D}

\def\be{\begin{equation}}
\def\ee{\end{equation}}
\def\bea{\begin{eqnarray}}
\def\eea{\end{eqnarray}}

\newcommand\firstchi{$\widetilde{\chi}_{1}^{0}$\xspace}

\def\secondchi{$\widetilde{\chi}_{2}^{0}$\xspace}
\def\thirdchi{$\widetilde{\chi}_{3}^{0}$\xspace}

\def\firstcharg{$\widetilde{\chi}_{1}^{\pm}$\xspace}

\def\chargneut{$\widetilde{\chi}_{1}^{\pm} \widetilde{\chi}_{2}^{0}$\xspace}
\def\firstchargpair{$\widetilde{\chi}_{1}^{\pm} \widetilde{\chi}_{1}^{\mp}$\xspace}
\def\chargneutthree{$\widetilde{\chi}_{1}^{\pm} \widetilde{\chi}_{3}^{0}$\xspace}
\def\neuttwoneutthree{$\widetilde{\chi}_{2}^{0} \widetilde{\chi}_{3}^{0}$\xspace}

\def\PW{W\xspace}
\def\PZ{Z\xspace}
\def\PH{H\xspace}
\def\WH{WH\xspace}
\def\WZ{WZ\xspace}
\def\WW{WW\xspace}
\def\HT{$H_{T}$\xspace}
\def\PV{V\xspace}
\def\ptmiss{$p^{miss}_{T}$\xspace}
\def\Pell{l\xspace}
\def\GeV{~\mathrm{GeV}\xspace}

\newcommand{\Photo}{\includegraphics[height=35mm]{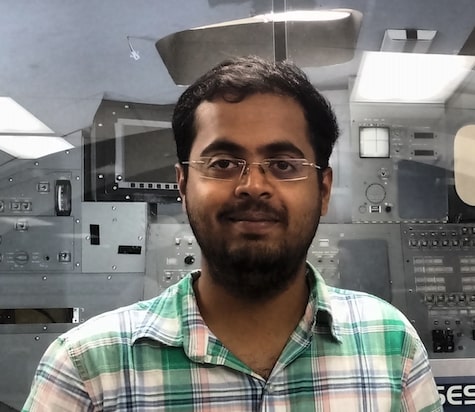}}

\begin{document}
\vspace*{4cm}
\title{SEARCH FOR ELECTROWEAK PRODUCTION OF CHARGINOS AND NEUTRALINOS IN ALL HADRONIC FINAL STATES AT THE CMS EXPERIMENT}

\author{ ANKUSH REDDY KANUGANTI \\ On behalf of the CMS Collaboration }

\address{Baylor University, Department of Physics, Waco, USA}

\maketitle\abstracts{
The results from a search for chargino-neutralino or chargino pair production via electroweak interactions are summarized. 
The results are based on a sample of $\sqrt{s}=13~\mathrm{TeV}$ proton-proton collisions from the LHC, recorded with 
the CMS detector~\cite{cms} and corresponding to an integrated luminosity of $137~\mathrm{fb}^{-1}$. The search considers final states 
with large missing transverse momentum and pairs of hadronically decaying bosons WW, WZ, and WH, which are identified using novel algorithms. 
No significant excess of events is observed relative to the expectation from the standard model. Limits at the $95\%$ confidence level are placed on 
the cross section for production of mass-degenerate wino-like superpartners of SU(2) gauge bosons, $\widetilde{\chi}_{1}^{\pm}$/$\widetilde{\chi}_{2}^{0}$. 
In the limit of nearly-massless neutralinos $\widetilde{\chi}_{1}^{0}$, $\widetilde{\chi}_{1}^{\pm}$ and $\widetilde{\chi}_{2}^{0}$ with masses up to 870 and 
$960~\mathrm{GeV}$ are excluded for $\widetilde{\chi}_{2}^{0}\to\mathrm{Z}\widetilde{\chi}_{1}^{0}$ and $\widetilde{\chi}_{2}^{0}\to\mathrm{H}\widetilde{\chi}_{1}^{0}$, 
respectively. Interpretations for other models are also presented.}

\section{Introduction}

Supersymmetry~\cite{su} (SUSY) proposes the addition of a new symmetry to the standard model (SM) of particle physics and proposes 
for each boson (fermion) in the SM, there is also a fermionic (bosonic) superpartner (sparticle). 
The results presented here search for electroweak production of sparticles under the assumption that strongly-coupled
sparticles are too massive to be produced at the LHC. Assuming that the superpartners of the SM leptons, the
sleptons, are much heavier than the charginos and neutralinos, the decays of charginos and neutralinos proceed through W, Z 
and Higgs (H) bosons. Using simplified models~\cite{smodel} of \chargneut and \firstchargpair production,
where the \firstcharg always decays to the \PW boson and the \firstchi, \secondchi decays 100\% of the time to either 
a \PZ or H plus the \firstchi. 
In \chargneut production, the \firstcharg and \secondchi are considered to be the wino-like mass-degenerate next-to-lightest supersymmetric particles (NLSPs),
while in \firstchargpair production, the \firstcharg is the wino-like NLSP.
Assuming that $R$-parity is conserved and that the \firstchi is a bino-like lightest supersymmetric particle (LSP) which escapes the detector unobserved.
Targeted final states are \WH, \WZ, or \WW together with a large transverse momentum imbalance.
The corresponding simplified models are referred to as TChiWH, TChiWZ, and TChiWW respectively.

\begin{figure}[h!]
       \centering
       \includegraphics[width=0.48\textwidth]{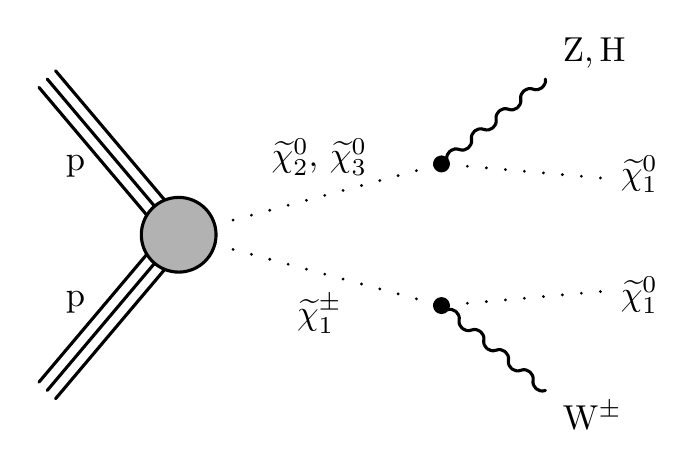}
       \includegraphics[width=0.48\textwidth]{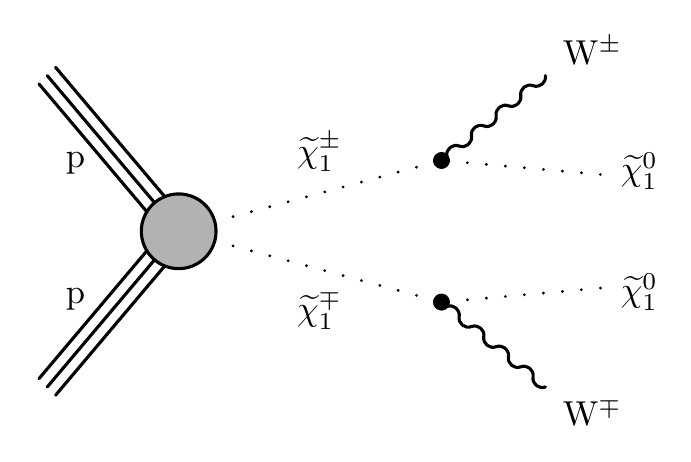}
       \caption{Production of \chargneut and \firstchargpair with the \firstcharg decaying to a \PW boson and \firstchi and the \secondchi decaying to 
       either a \PZ boson or a Higgs boson and \firstchi [4].}
       \label{fig:model}
\end{figure}
       
\section{Event Selection}
All events in the four signal regions (SRs) are required to pass a common set of baseline selection criteria.  Each event is required to have a primary 
vertex and no isolated leptons, photons, or isolated tracks . Few other selections include \ptmiss $> 200\GeV$ and $H_{T} > 300\GeV$. 
Large \ptmiss and \HT are typical of chargino and neutralino production when a high-momentum boson is present. 
For signal events we require at least two AK8 jets and 2--6 (inclusive) AK4 jets. 
Within this baseline phase space, four SRs are defined.  Three SRs require at least one b-tagged AK4 jet ($n_{b}\geq1$), referred to as the b-tag regions.
The remaining SR requires zero b-tagged AK4 jets ($n_{b}=0$), referred to as the b-veto region.
In addition to these SRs, there are several control regions (CRs), which are used to help constrain the background estimates.

\subsection{The b-veto search region}
The b-veto SR seeks to isolate events that are consistent with the production of \WW, \WZ pairs of bosons plus large \ptmiss.  
In addition to the baseline event selection described above, the b-veto SR requires that at least two AK8 jets satisfy $65 < m_{J} < 105\GeV$.
At least one AK8 jet must be \PW tagged, and at least one other AK8 jet must be \PV tagged.

\subsection{The b-tag search region}
The main b-tag SR which is most sensitive is the WH signal region. The WH SR requires at least one W boson candidate 
$\Delta R(\text{b\textendash jet}, \text{AK8 jet}) < 0.8$  with the AK8 jet mass $65 < m_{J} < 105\GeV$ and W-tagged and at least one Higgs 
boson candidate $\Delta R(\text{b\textendash jet}, \text{AK8 jet}) > 0.8$  with the AK8 jet mass $75 < m_{J} < 140\GeV$ that is bb tagged.

\section{Background Estimation}
For the b-veto SR, the background yields in the b{}-veto SR are estimated using two sets of transfer factors derived from simulation, $\mathcal{R}_{i}$,
defined as the ratio of the summed 0- and 1-res event yields in the
SR with respect to either the zero-tag or one-tag CR.
The values of $\mathcal{R}_{i}$ are computed separately for each \ptmiss
bin and typically range between 0.2 and 0.3.
The contributions of rare processes to the SRs and CRs are taken from simulation with appropriate data-to-simulation corrections applied.
The total background prediction is given by:
\be
N^\text{data}_\text{SR} = \mathcal{R}_{i} ( N^\text{data}_{\text{CR}_i} - N^\text{MC}_{\text{CR}_i,\text{rare}} )+ N^\text{MC}_\text{SR,rare}\
\label{eq:bveto_pred}
\ee
where $\mathcal{R}_{i} = N^\text{MC}_\text{SR,0\&1-res} / N^\text{MC}_{\text{CR}_i,\text{0\&1-res}}$
and $\text{CR}_i$ is either the zero-tag or one-tag CR. The final background predictions for the SR are determined by a simultaneous fit of the two CRs.
\\
\\
For the WH SR, the background yield is estimated for top and non-resonant processes separately. 
A transfer factor, $\mathcal{R}_{0\Pell/1\Pell}$, is used to provide an
estimate of the number of top background events in either the SR or the $0\Pell$ antitag CR. The
values of $\mathcal{R}_{0\Pell/1\Pell}$ are computed from simulation, including all
corrections to the lepton reconstruction efficiencies, b-tagging efficiencies, and AK8 jet tagging efficiencies.
The predicted number of top background events in either the SR or the $0\Pell$ antitag CR is given by:
\be
\label{eq:reson_pred}
N^{\text{pred},0\Pell}_{i,\text{top}} = \frac{N^{\text{MC},0\Pell}_{i,\text{top}}}{N^{\text{MC},1\Pell}_{i,\text{all}}} N^{\text{data},1\Pell}_{i} = \mathcal{R}_{0\Pell/1\Pell}N^{\text{data},1\Pell}_{i}
\ee
where $N^\text{MC}$ denotes the number of events expected from simulation, $N^\text{data}$ denotes the number of observed events, and $N^\text{pred}$ denotes the number of events predicted via this method.
Additionally, the subscript $i$ denotes the tagging region, tag or antitag.
The subscript ``all'' refers to all of the SM backgrounds, while ``top'' refers to only the top background.
\\
Using $\mathcal{R}_{\text{p/f}}$ and the prediction of top backgrounds described above,
the predicted 0-res background contribution to the SR is given by:
\be
\label{eq:nonreson_pred}
N^{\text{pred},0\Pell}_{\text{0-res}} = \mathcal{R}_\text{p/f}\left(N^{\text{data},0\Pell}_{\text{antitag}}-N^{\text{pred},0\Pell}_{\text{antitag,top}}-N^{\text{MC},0\Pell}_{\text{antitag,rare}}\right)
\ee
where $N^{\text{data},0\Pell}_{\text{antitag}}$ denotes the number of observed events in the $0\Pell$ antitag CR,
$N^{\text{pred},0\Pell}_{\text{antitag,top}}$ denotes the predicted number of top background events from Eq.~(\ref{eq:reson_pred}), and
$N^{\text{MC},0\Pell}_{\text{antitag,rare}}$ denotes the number of rare background events, such as diboson and triboson events, expected from simulation.
\\
\section{Results}
Fits to the SRs and CRs are performed using a statistical model of our SM background predictions.
This fitting procedure further constrains the predictions and the uncertainties in the predictions.
The predicted SM backgrounds based on this
procedure, the observations, and the predicted signal yields in each of the SRs are shown below.
No statistically significant excess of events is observed in the data with respect to the SM background predictions.

\begin{figure}[h!]
       \centering
       \includegraphics[width=0.48\textwidth]{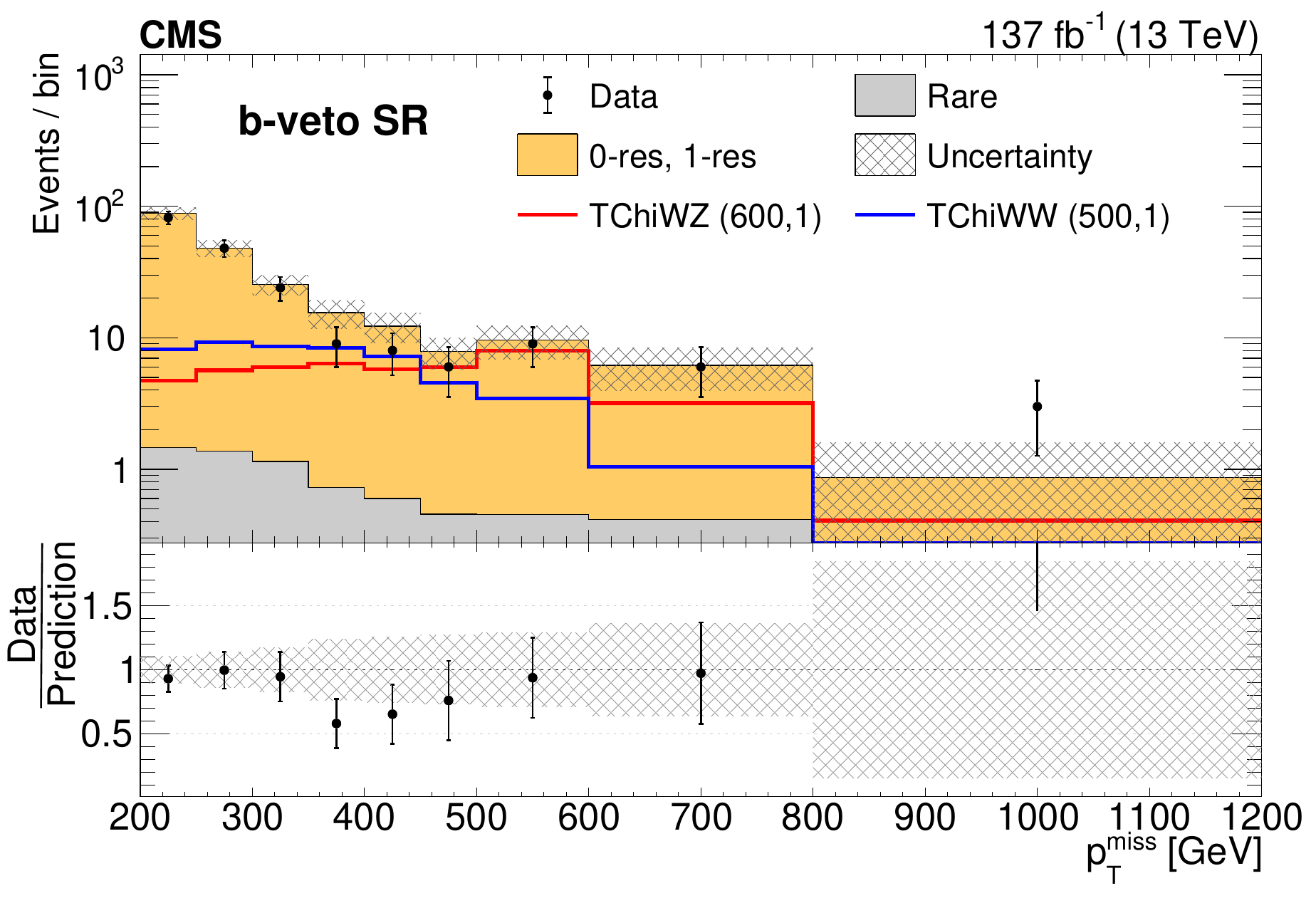}
       \includegraphics[width=0.48\textwidth]{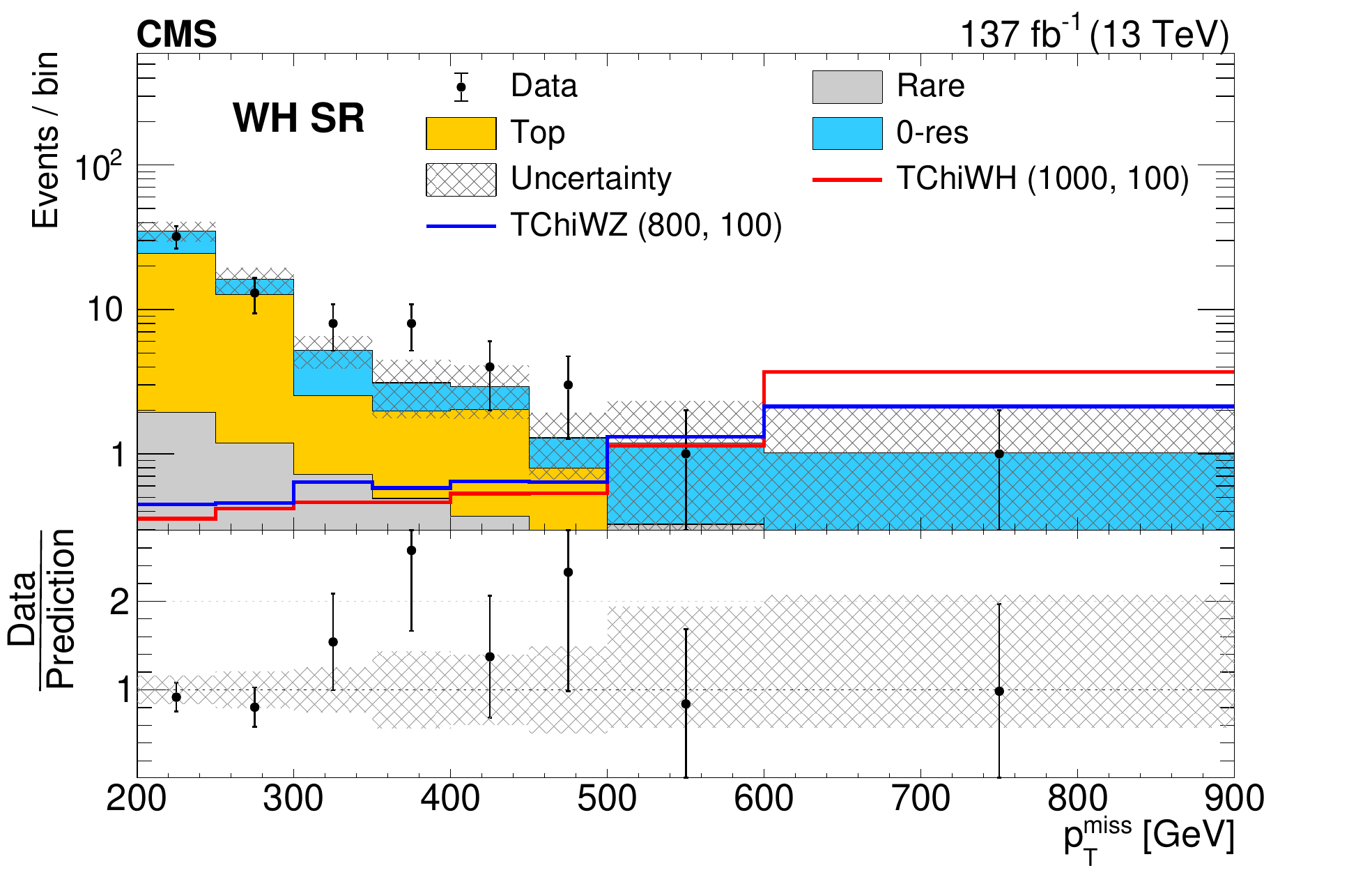}
       \caption{Prediction vs.\ data in the b-veto SR (left), the \WH SR (right).
       The filled histograms show the SM background predictions, and the
       open histograms show the expectations for selected signal models, which are
       denoted in the legend by the name of the model followed by the assumed masses
       of the NLSP and LSP. The observed event yields are indicated by black markers [4].}
      \label{fig:metplot}
\end{figure}

\section{Summary}
Using wino-like pair production cross sections, 95\% confidence level (CL) mass exclusions are derived.
For signals with \WW, \WZ, or \WH final states,
the NLSP mass exclusion limit for low-mass LSPs extends up to 670, 760, and 970$\GeV$, respectively.
When we consider models including both wino-like NLSP \chargneut and \firstchargpair production with
either \secondchi$\to$\PZ\firstchi or \secondchi$\to$\PH\firstchi,
the NLSP mass exclusion extends up to 870 and 960$\GeV$, respectively.
These mass exclusions are the most stringent constraints to date set by CMS at high NLSP masses. 

\begin{figure}[h!]
       \centering
       \includegraphics[width=0.32\textwidth]{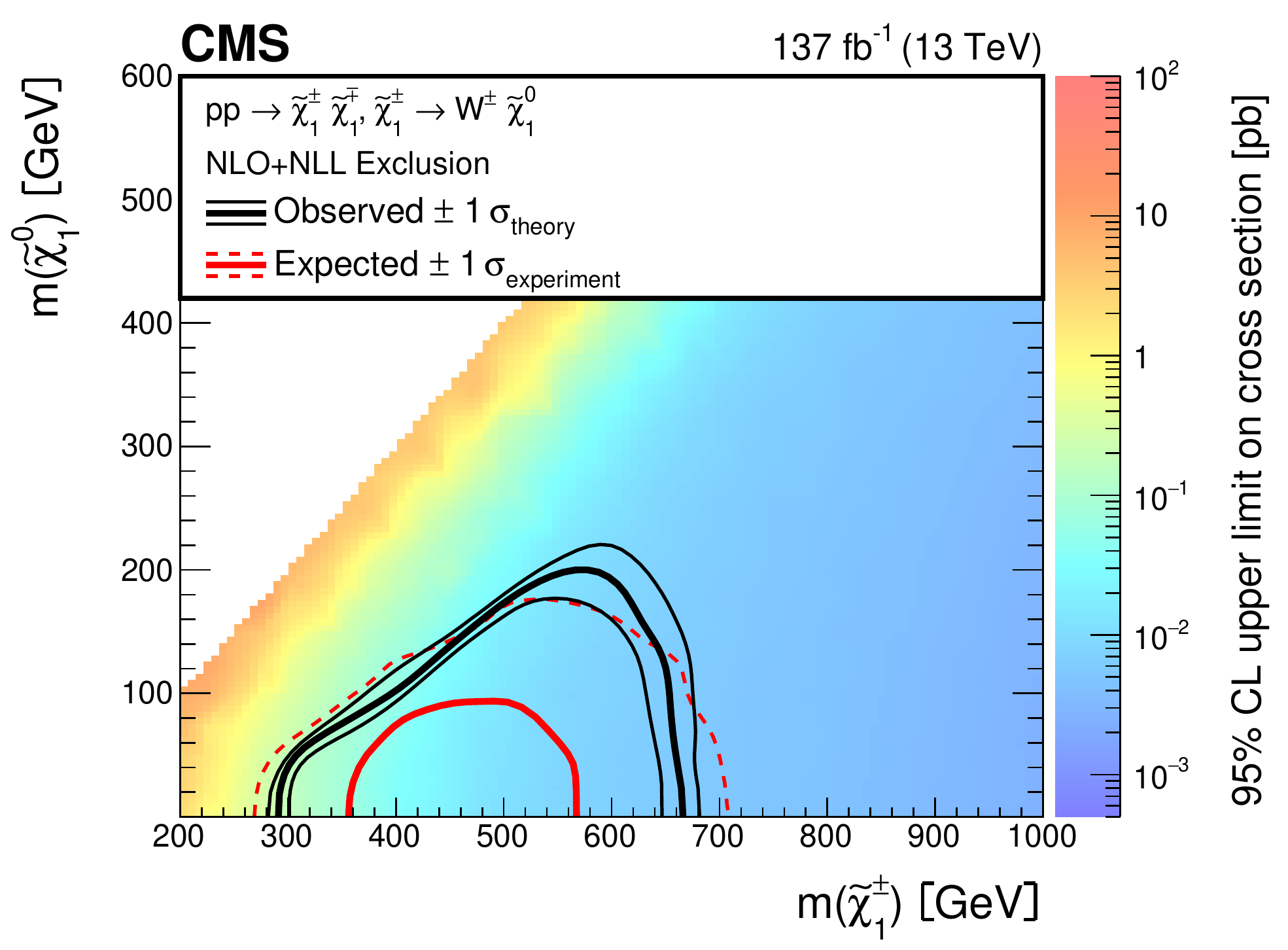}
       \includegraphics[width=0.32\textwidth]{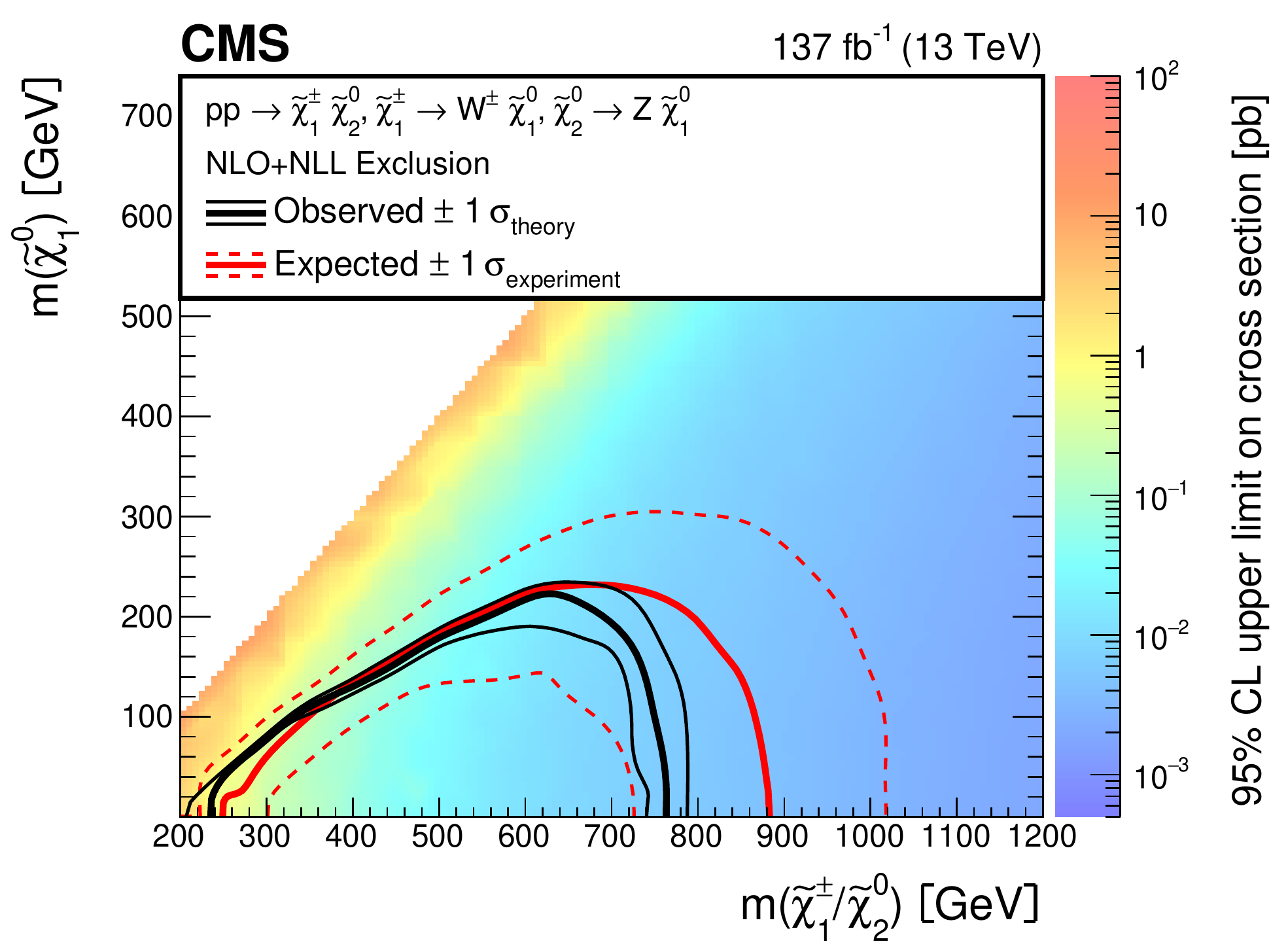}
       \includegraphics[width=0.32\textwidth]{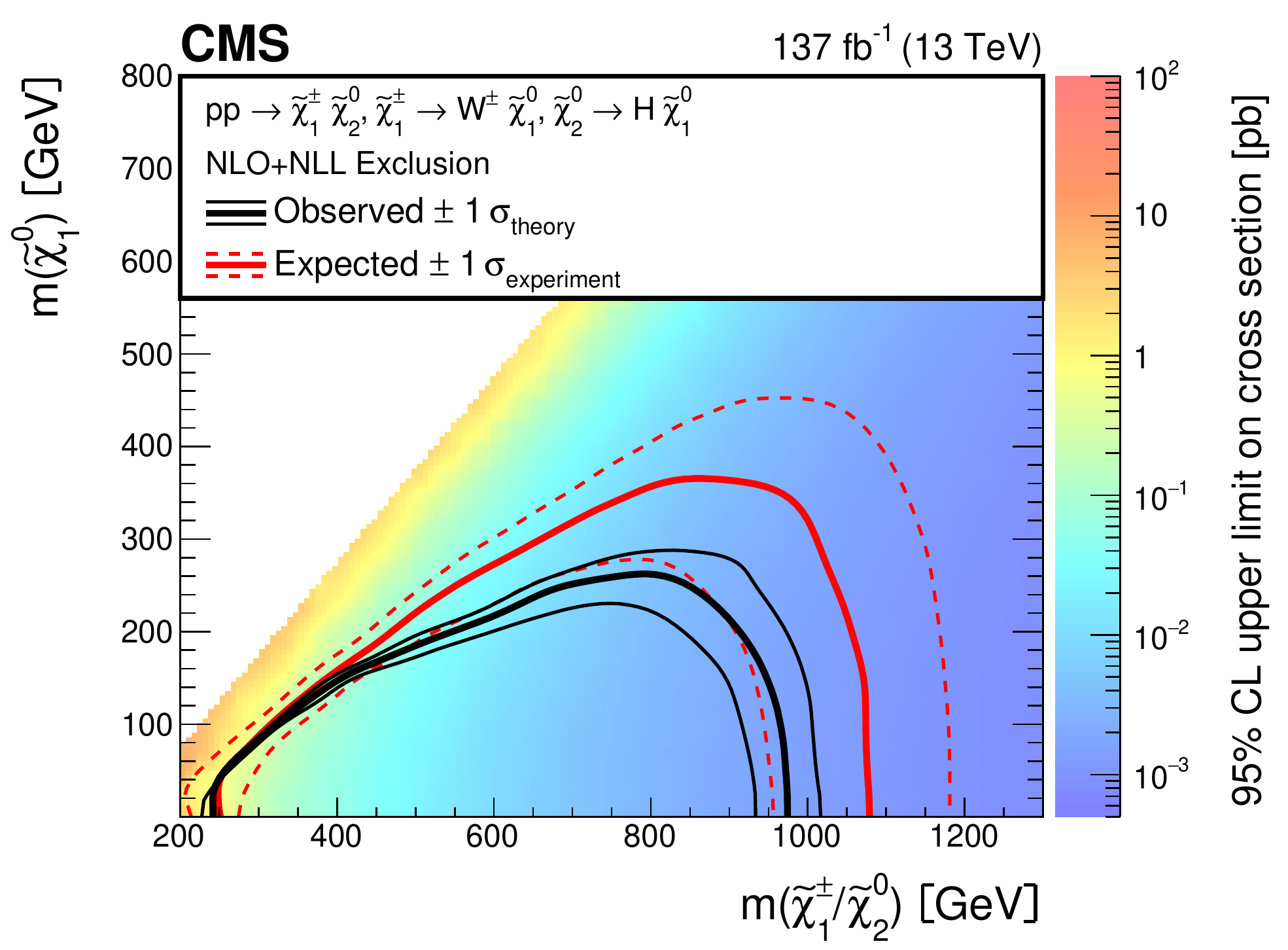}
	\caption{The 95\% CL upper limits on
	the production cross sections for \firstchargpair assuming
	that each \firstcharg decays to a \PW boson and \firstchi (left) and \chargneut production assuming
	that the \firstcharg decays to a \PW boson and \firstchi and that the \secondchi decays to a \PZ boson and \firstchi (middle)
	or that the \secondchi decays to a \PH and \firstchi (right). The black curves represent 
       the observed exclusion contour and the change in this contour due to variation of these
       cross sections within their theoretical uncertainties ($\sigma_{\text{theory}}$).
       The red curves indicate the mean expected exclusion contour
       and the region containing 68\% ($\pm 1\,\sigma_{\text{experiment}}$) of the expected
       exclusion limits under the background-only hypothesis [4].
       } 
       %The mass exclusion limits are computed assuming wino-like cross sections.
       \label{fig:limits}
\end{figure}

Results are also shown using the higgsino-like NLSPs \firstcharg, \secondchi, and \thirdchi.
NLSP masses between 300 and 650$\GeV$ are excluded for low mass LSPs at 95\% CL under the standard model hypothesis;
however, the observed cross section upper limits lie mostly below the theoretical cross section because of a modest excess in data.

%for review process 
%\footnote{Figure 4 has been approved, shown in the conference and the paper which has this figure is ready for submission very soon}

\begin{figure}[h!]
       \centering
       \includegraphics[width=0.48\textwidth]{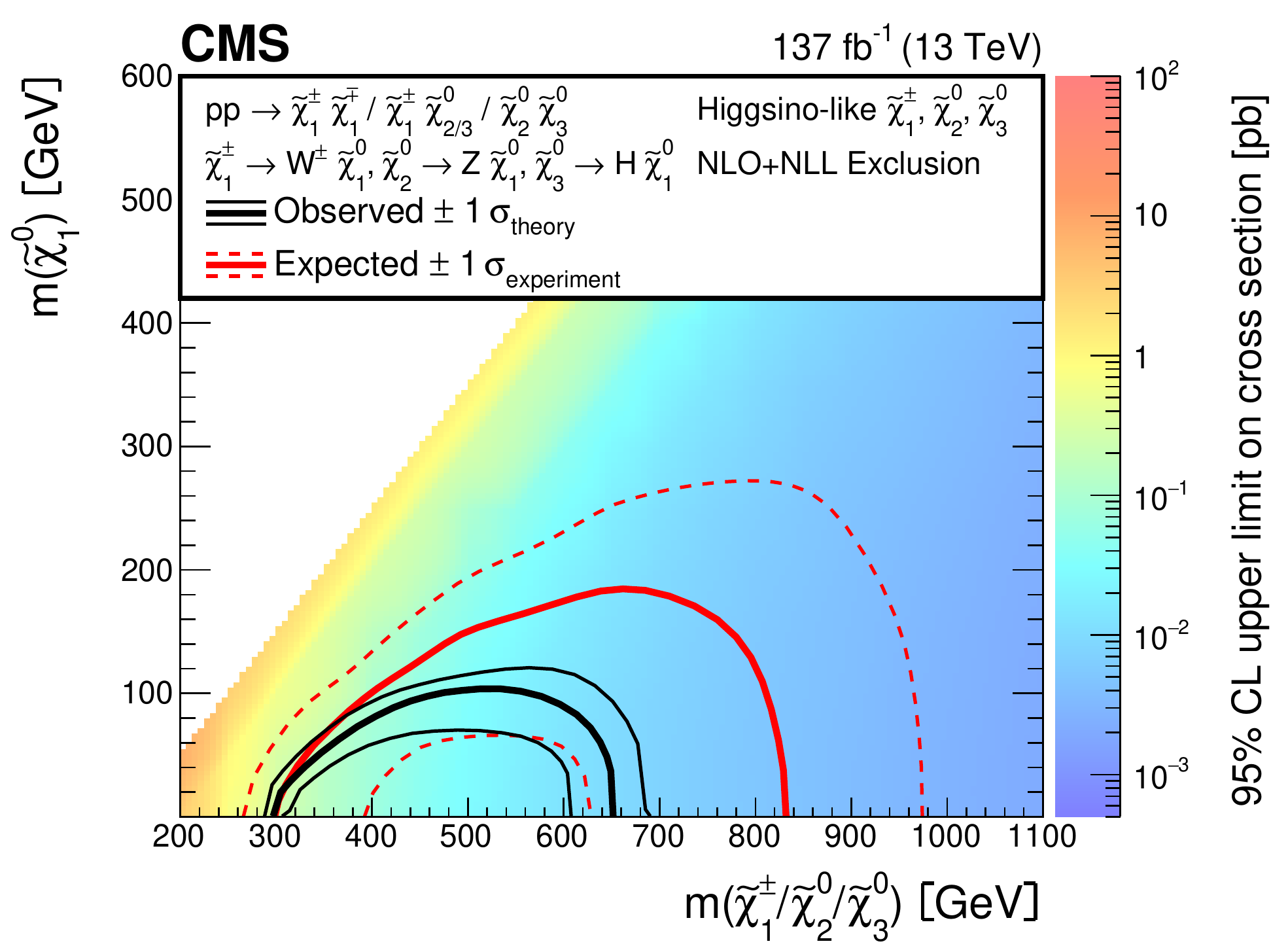}
       \caption{Expected and observed 95\% CL exclusion for mass-degenerate higgsino-like \firstchargpair, \chargneut, 
       \chargneutthree, and \neuttwoneutthree production as functions of the NLSP and LSP masses. The 95\% CL upper 
       limits on the production cross sections are also shown. The \firstcharg, \secondchi, and \thirdchi are considered to be mass degenerate [4].}
       \label{fig:limits3}
\end{figure}

\end{document}